\documentclass[12pt]{article}

\usepackage{graphicx}
\usepackage{scicite}
\usepackage{aas_macros}
\usepackage{times}

\newenvironment{sciabstract}{%
\begin{quote} \bf}
{\end{quote}}

\newcounter{lastnote}
\newenvironment{scilastnote}{%
\setcounter{lastnote}{\value{enumiv}}%
\addtocounter{lastnote}{+1}%
\begin{list}%
{\arabic{lastnote}.}
{\setlength{\leftmargin}{.22in}}
{\setlength{\labelsep}{.5em}}}
{\end{list}}

\def\gtrsim{\mathrel{\hbox{\rlap{\hbox{\lower4pt\hbox{$\sim$}}}\hbox{$>$}}}}

\title{Major Galaxy Mergers and the Growth of Supermassive Black Holes in Quasars} 

\author
{Ezequiel Treister,$^{1,2\ast}$ Priyamvada Natarajan,$^{3,4,5\ast}$ David B. Sanders,$^{1}$ \\
C. Megan Urry,$^{3,4,5}$ Kevin Schawinski,$^{2,4,5}$ Jeyhan Kartaltepe.$^{1,6}$\\
\\
\normalsize{$^{1}$Institute for Astronomy, 2680 Woodlawn Drive, University of Hawaii, Honolulu, HI 96822}\\
\normalsize{$^{2}$Chandra/Einstein Fellow.}\\
\normalsize{$^{3}$Department of Astronomy, Yale University, PO Box 208101, New Haven, CT 06520}\\
\normalsize{$^{4}$Department of Physics, Yale University, P.O. Box 208121, New Haven, CT 06520}\\
\normalsize{$^{5}$Yale Center for Astronomy and Astrophysics, P.O. Box 208121, New Haven, CT 06520}\\
\normalsize{$^{6}$NOAO, 950 N. Cherry Avenue, Tucson, AZ 85719}\\
\\
\normalsize{$^\ast$To whom correspondence should be addressed;}\\
\normalsize{E-mail:  treister@ifa.hawaii.edu, priyamvada.natarajan@yale.edu.}
}

\date{}

\begin{document}

\baselineskip24pt

\maketitle

\begin{sciabstract}
Despite observed strong correlations between central supermassive
black holes (SMBHs) and star-formation in galactic nuclei,
uncertainties exist in our understanding of their coupling. We present
observations of the ratio of heavily-obscured to unobscured quasars as
a function of cosmic epoch up to $z$$\simeq$3, and show that a simple
physical model describing mergers of massive, gas-rich galaxies
 matches these observations. In the context of
this model, every obscured and unobscured quasar represent two
distinct phases that result from a massive galaxy merger event. Much
of the mass growth of the SMBH occurs during the heavily-obscured
phase. These observations provide additional evidence for a causal link
between gas-rich galaxy mergers, accretion onto the nuclear SMBH and
coeval star formation.
\end{sciabstract}

While unobscured quasars \cite{note1} have been known for a long time
\cite{schmidt63} and their statistical properties are well-studied
\cite{richards09}, the numbers of the heavily obscured quasar
population and its variation with cosmic epoch are still strongly
debated. This population has been uncovered using multi-wavelength
selection techniques that simultaneously exploit X-ray
\cite{barger03}, optical \cite{zakamska03} and mid-infrared
\cite{martinez06} wavelengths. As a result of the efficiency of these
techniques, the sample sizes of obscured quasars is growing
substantially. The existence of a large number of heavily obscured
quasars at $z$$\sim$2 was predicted by early AGN population synthesis
models that successfully explain the generation of the X-ray
background \cite{gilli01}.  However, their space density cannot be
constrained by these calculations \cite{treister09b}. What population
these obscured sources evolve into or proceed from is poorly
understood at present. 

The link between Ultra-luminous Infrared Galaxies (ULIRGs) and quasars
was first suggested by Sanders et al. \cite{sanders88}. There is
substantial observational evidence that ULIRGs, at least locally, are
the product of the gas-rich merger of two massive
($M$$\gtrsim$10$^{11}$~$M$$_\odot$) galaxies \cite{sanders96}. The merger process is
believed to switch on accretion onto the central black hole as it
provides efficient transport of gas to the nucleus \cite{barnes91}.
The gas funneled to the center is expected to fuel the supermassive
black hole and induce star formation. The origin of the infrared
luminosity of these sources, whether they are powered primarily by
star formation processes \cite{genzel98} or the AGN \cite{sanders88}
activity, is still debated. Here, we present recent measurements of
the space density of heavily-obscured quasars as a function of
redshift, and estimate the duration of the obscured stage 
by comparing models with observations.

Mainly due to the effects of dust and gas obscuration at most wavelengths, finding
heavily obscured quasars \cite{note2} is a challenging task which has
prevented the identification of larger samples, in particular, at high
redshifts. Measuring their space density is even more difficult, as it
requires a good knowledge of the selection function and observational
biases. We have compiled observations of obscured quasars selected
at various wavelengths using different techniques, described in detail
in the SOM, including spectral fitting in X-rays \cite{tozzi06} and infrared
selection \cite{alexander08,fiore09,treister09c}. X-rays, specially at
rest-frame energies greater than 10 keV, are not significantly
affected by obscuration. In addition, most of the absorbed energy is later 
re-emitted at infrared wavelengths. Thus, these techniques permit an 
estimate of the number of heavily-obscured quasars at $z$$>$1. In the 
local Universe, the space density of ULIRGs \cite{kim98}, which can be used
as an indicator of the total number of quasars, implies that the ratio of
heavily-obscured to unobscured quasars at $z$=0.1 is $\sim$1 (Fig. 1). 

Both in the local Universe and at $z$$\simeq$1, heavily-obscured AGN
are bright at mid- and far-infrared wavelengths and also potentially in the 
sub-millimeter but are optically faint (Fig. 2). At low redshift, an ULIRG is 
clearly the product of a galaxy merger, while at high redshift a merger is 
suggested, but deeper data are needed to confirm this hypothesis. The 
rest-frame optical images of six heavily-obscured quasar candidates
show indications of ongoing major mergers and interactions (Fig. 3).

In order to interpret these observations and the evolution of these
populations, we started with the standard ansatz that the gas-rich major
merger of two massive galaxies produces one newly-fueled quasar. This
triggered quasar is originally obscured by the surrounding gas and
dust \cite{sanders88}, in some cases reaching Compton-thick
levels. (i.e., where the optical depth for Compton scattering is greater
than one.) After a time $\Delta$t $\sim$10$^7$-10$^8$ years
\cite{hopkins08}, which we estimate independently below, most of the
dust and gas are removed from the central region and the quasar
becomes unobscured.  

To test this simple prescription, the calculated ratio of heavily obscured
to unobscured sources from merger rates of massive gas-rich galaxies
needs to match the observed ratio of obscured to unobscured
quasars. This ratio can be calculated using the merger rate as a
function of cosmic time in the context of the hierarchical cold dark
matter (LCDM) structure formation paradigm. Using the assumptions
described above, we estimated the ratio of obscured to unobscured
sources as:

\begin{equation}
\frac{N_{\rm obsc}}{N_{\rm unobsc}}(z)=\frac{\Delta t \frac{d^2 {\rm Merger}}{dNdt} N_{\rm gal}(>M_{\rm star}(z))f_{\rm g}(z)}{N_{\rm unobsc}(z)},
\end{equation}

where $N_{\rm obsc}$ is the space density of heavily-obscured quasars;
$d^2$Merger/$dNdt$ is the merger frequency per galaxy per unit time;
$N_{\rm gal}$ and $N_{\rm unobsc}$ are the space densities of massive
galaxies and unobscured quasars, respectively, and $f_{\rm g}$ is the
average fraction of gas-rich galaxies. The major merger frequency per
galaxy in the LCDM paradigm can be parameterized as a power-law in
(1+$z$) with a mass-dependent exponent of $\simeq$1.5
\cite{hopkins09}. This form is derived from model parameters
constrained by observations. In order to estimate the space density of
galaxies above a threshold stellar mass, we used the median mass
measured for ULIRGs found in the Cosmic Evolution (COSMOS) survey. The
median mass increases with decreasing redshift, going from
$\sim$10$^{11}$M$_\odot$ at $z$$\sim$2 to 10$^{11.3}$M$_\odot$ at
$z$=0.8 (Fig. S1). We then incorporated this limiting mass into the stellar 
mass function computed by Marchesini et al. \cite{marchesini09} to obtain 
the space density of massive galaxies as a function of redshift. Rather than 
directly estimate the gas content of high redshift galaxies, which is currently
observationally impossible at these redshifts, we used the average star
formation rate as a proxy for the evolution of the fraction $f_{\rm
g}$ of gas-rich galaxies. The evolution of the star formation rate can
be approximated as (1+$z$)$^{2}$ up to $z$$\simeq$2, remaining mostly
flat at higher redshifts, based on ultraviolet observations of
galaxies up to $z$$\simeq$2.5 \cite{dahlen07}. Finally, the space
density of unobscured quasars, $N_{\rm unobsc}$($z$) has been measured
by both X-ray \cite{hasinger05,barger05} and optical
\cite{richards06b} surveys, and consistent results are found with
these two methods. These are all the ingredients needed to compute the
expected fraction of obscured to unobscured quasars, wherein $\Delta$t
in equation (1) can be determined as a free parameter. The redshift
dependence of each of these components is shown in Fig.~S2.

The estimates from our simple scenario are consistent with the
observations (Fig. 1), in particular considering the steep
evolution in the relative number of obscured sources from $z$=1.5 to
3. This rapid increase is not predicted or expected from existing AGN
luminosity functions (Fig. 1). That is, extrapolations from the behavior of 
less-obscured lower-luminosity sources (with observed column densities 
$<$10$^{23}\,$cm$^{-2}$) do not match current observations, in particular 
at $z$$>$2. This suggests the existence of a different channel for the triggering 
mechanism for quasars, compared to lower-luminosity AGN. The best-fit value of
$\Delta$t we obtained is 96$\pm$23 Myrs (90\% confidence level). This is very similar 
to the current best estimates of quasar lifetimes in their optically bright, unobscured
phases [10-100 Myrs \cite{martinez09}], indicating that these sources
spend roughly half their life in the obscured phase.

Having determined that quasars that are fueled by the merger of
massive gas-rich galaxies spend comparable amounts of time in the the
obscured and unobscured phases, we proceeded to estimate the
implications for the mass accretion onto the nuclear supermassive
black holes during these two stages. Assuming a typical accretion
efficiency of $\varepsilon$=0.1, the mass growth of the supermassive
black hole due to accretion is given by:

\begin{equation}
\Delta M_{\rm BH}=1.6\times 10^7 \left( \frac{L_{\rm bol}}{10^{45}{\rm erg/s}}\right)\left(\frac{T}{10^{8} {\rm yrs}}\right)M_\odot,
\end{equation}

where $T$ is the duration of the entire accretion episode that
includes the obscured and unobscured phases. The typical bolometric
luminosities of these sources span the range 10$^{45}$-10$^{47}$~erg/s
($\sim$10$^{12}$-10$^{14}$~L$_\odot$), while they have black hole
masses of M$_{\rm BH}$$\sim$10$^8$-10$^{10}$M$_\odot$
\cite{dietrich09,vestergaard09}. Given that the typical duration of
the total (both obscured and unobscured) luminous quasar phase is
$T$$\sim$2$\times$10$^8$ years, it is possible for a quasar to build
most or all of the black hole mass in a single event, which is
triggered by a major merger as suggested here.

In spite of copious accretion, due to their high redshifts and strong
absorption, this unexpected population of obscured quasars are not
key contributors to the extragalactic X-ray background
radiation. Their total contribution is estimated to be $\sim$1-2\%
\cite{treister09b}. However, they comprise a significant fraction of
the super-massive black hole mass density in the Universe \cite{haehnelt98}. 
Adding the extra obscured accretion reported here, which lasts as long as the optically bright phase,
increases our original estimate of the integrated black hole mass
density at $z$=0, by 4\%, to 4.5$\times$10$^5$ $M$$_\odot$Mpc$^{-3}$
\cite{treister09b}. Including this additional contribution, the
integrated black hole growth in the obscured quasar phase is
1.3$\times$10$^{5}$$M$$_\odot$Mpc$^{-3}$, or $\sim$30\% of the total
black hole mass density at $z$=0, in agreement with our conclusion
that the obscured quasar phase can harbor a large fraction of the
black hole growth\cite{note3}. Our results are in agreement with
recent estimates \cite{martinez09} that suggest an average accretion
efficiency of $\leq$10\% even accounting for heavily obscured
accretion.

\begin{scilastnote}
\item Support for the work of ET and KS was provided by the National
Aeronautics and Space Administration through Chandra/Einstein
Post-doctoral Fellowship Award Numbers PF8-90055 and PF9-00069
respectively issued by the Chandra X-ray Observatory Center, which is
operated by the Smithsonian Astrophysical Observatory for and on
behalf of the National Aeronautics Space Administration under contract
NAS8-03060. PN wishes to acknowledge the Radcliffe Institute for
Advanced Study where this work was started. CMU acknowledges 
support from NSF grant AST-0407295.
\end{scilastnote}

\clearpage

\begin{figure*}
\centering
\includegraphics[scale=0.7]{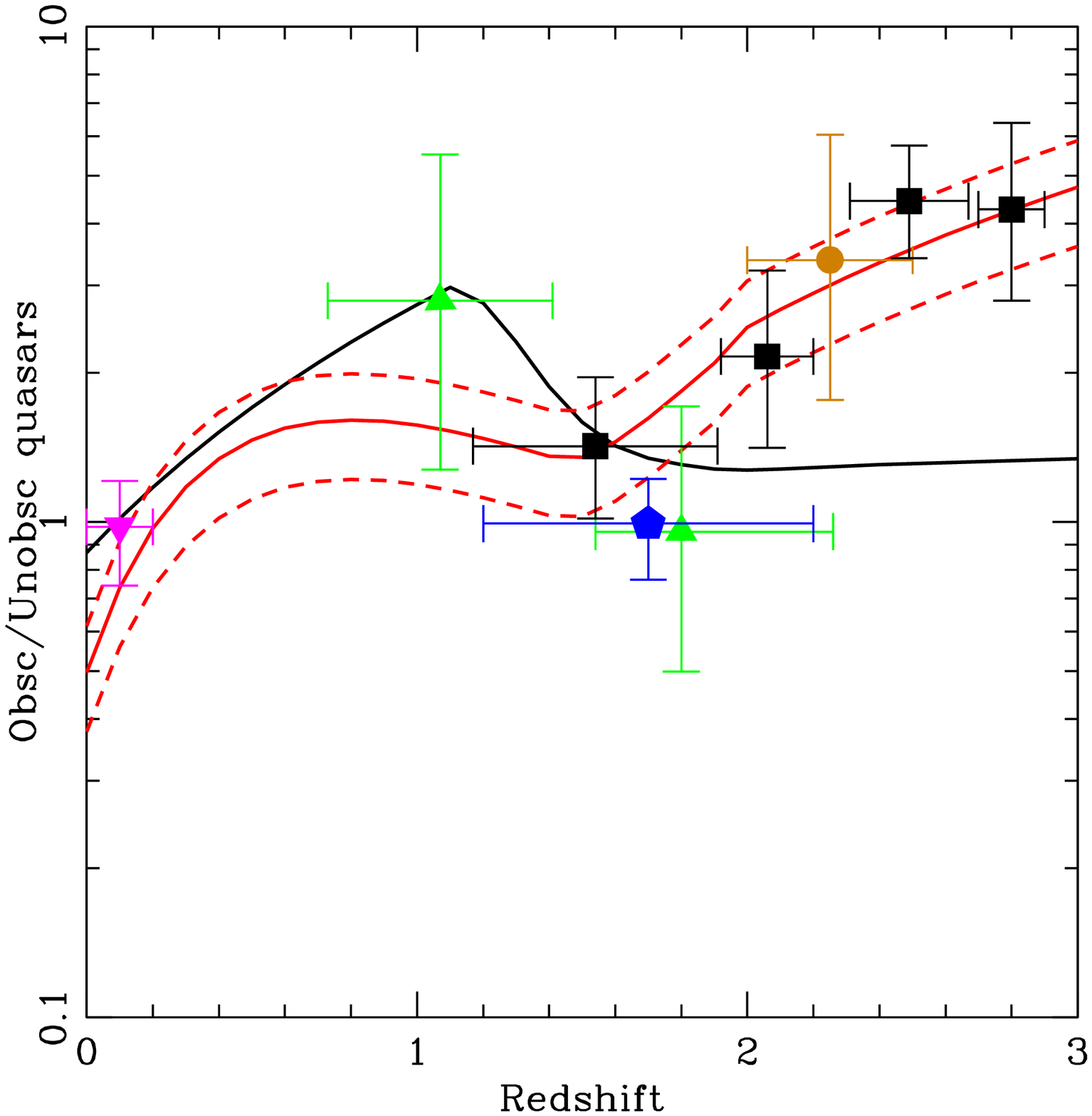}
\noindent{\\{\bf Fig.~1.} The ratio of heavily-obscured to unobscured
quasars as a function of redshift. Measurements of the space density
of obscured quasars at high redshift were obtained from X-ray --- {\it
green triangles} \cite{tozzi06} --- and mid-IR imaging --- {\it blue
pentagon} \cite{fiore09} and {\it black squares} \cite{treister09c}
--- and spectroscopy --- {\it brown circle} \cite{alexander08} ---
selection techniques. For the $z$$\simeq$0 measurement we used the
luminosity function of local ULIRGs \cite{kim98}, assuming that each
ULIRG is either a heavily-obscured or an unobscured quasar.  The {\it
solid black line} shows the heavily-obscured to unobscured quasar
ratio expected from AGN luminosity functions derived from hard X-ray
observations \cite{dellaceca08a}, while the {\it red solid line}
corresponds to the ratio obtained if every gas-rich major merger of two
massive galaxies generates a heavily-obscured quasar, which
after a time $\Delta$t$\simeq$96 Myrs becomes unobscured. {\it Dashed lines} show the uncertainty in this relation,
at the 90\% confidence level. }
\end{figure*}

\vfill
\eject

\begin{figure*}
\centering
\includegraphics[scale=0.65]{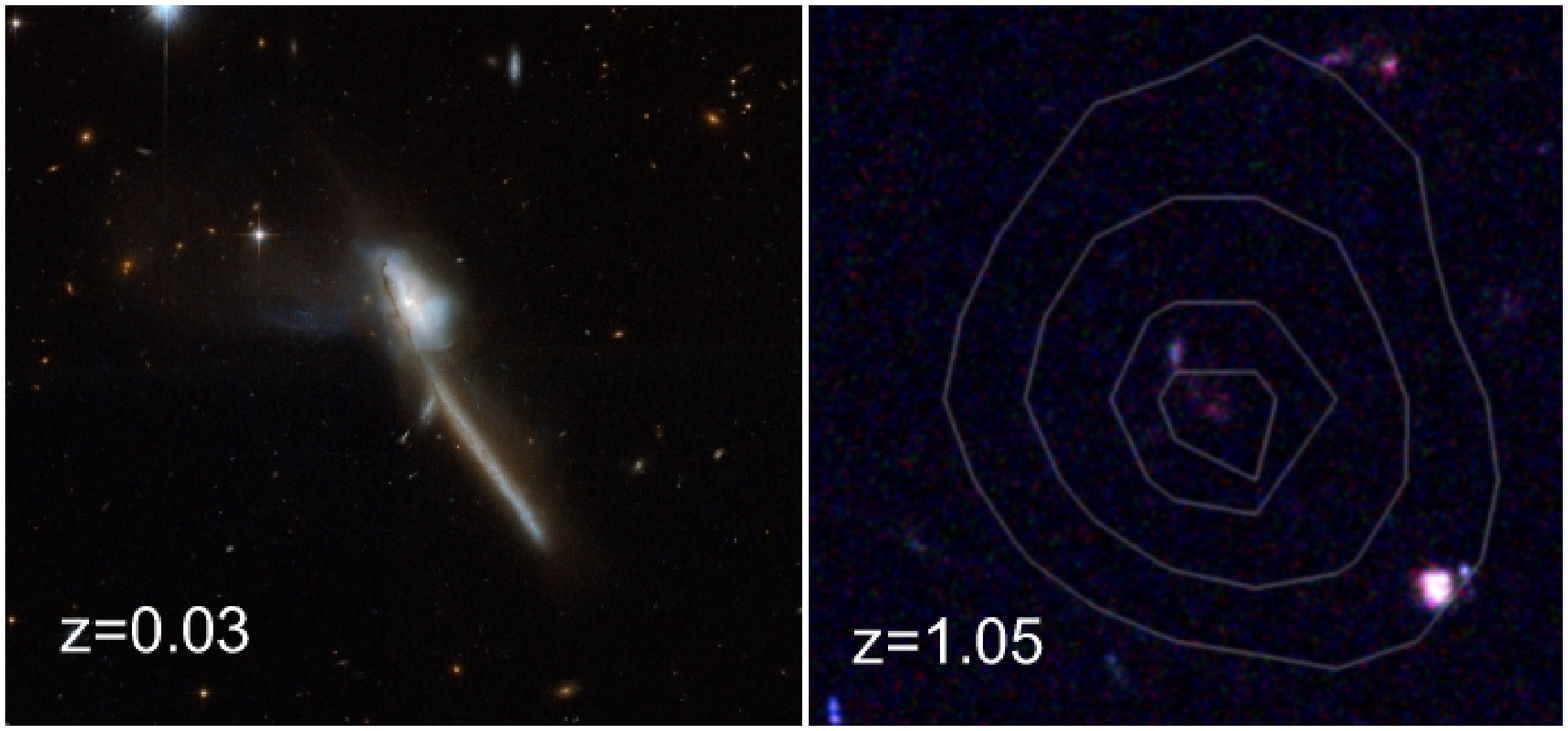}
\noindent{\\{\bf Fig.~2.} Optical images of two examples of
heavily-obscured, luminous AGN observed with the Hubble Space
Telescope. Filters used are F435W (B) and F814W (I) and V,I,z in the left and right panels 
respectively, while the physical size of both images is $\sim$90 kpc$\times$90 kpc. Contours in the right panel show
the Spitzer emission. The left panel shows the $z$=0.03 prototype ULIRG Mrk 273,
while in the right panel we present a high redshift heavily-obscured
quasar candidate in the GOODS-South field, with no X-ray
detection. The former is part of the IRAS 1 Jy sample \cite{kim98} and
has clear indications of a major merger, while X-ray observations with
Suzaku reveal Compton-thick obscuration levels
($N_{\rm H}$$>$10$^{24}$~cm$^{-2}$) and intrinsic quasar-like luminosities
\cite{teng09}. The high-$z$ source was selected as a Compton-thick AGN
candidate based on its high mid-IR to optical flux ratio and red
optical/near-IR colors. Left panel credit: NASA, ESA, the Hubble
Heritage Team (STScI/AURA)-ESA/Hubble Collaboration and A. Evans
(University of Virginia, Charlottesville/NRAO/Stony Brook University).
}
\end{figure*}

\vfill
\eject

\begin{figure*}
\centering
\includegraphics[scale=0.25]{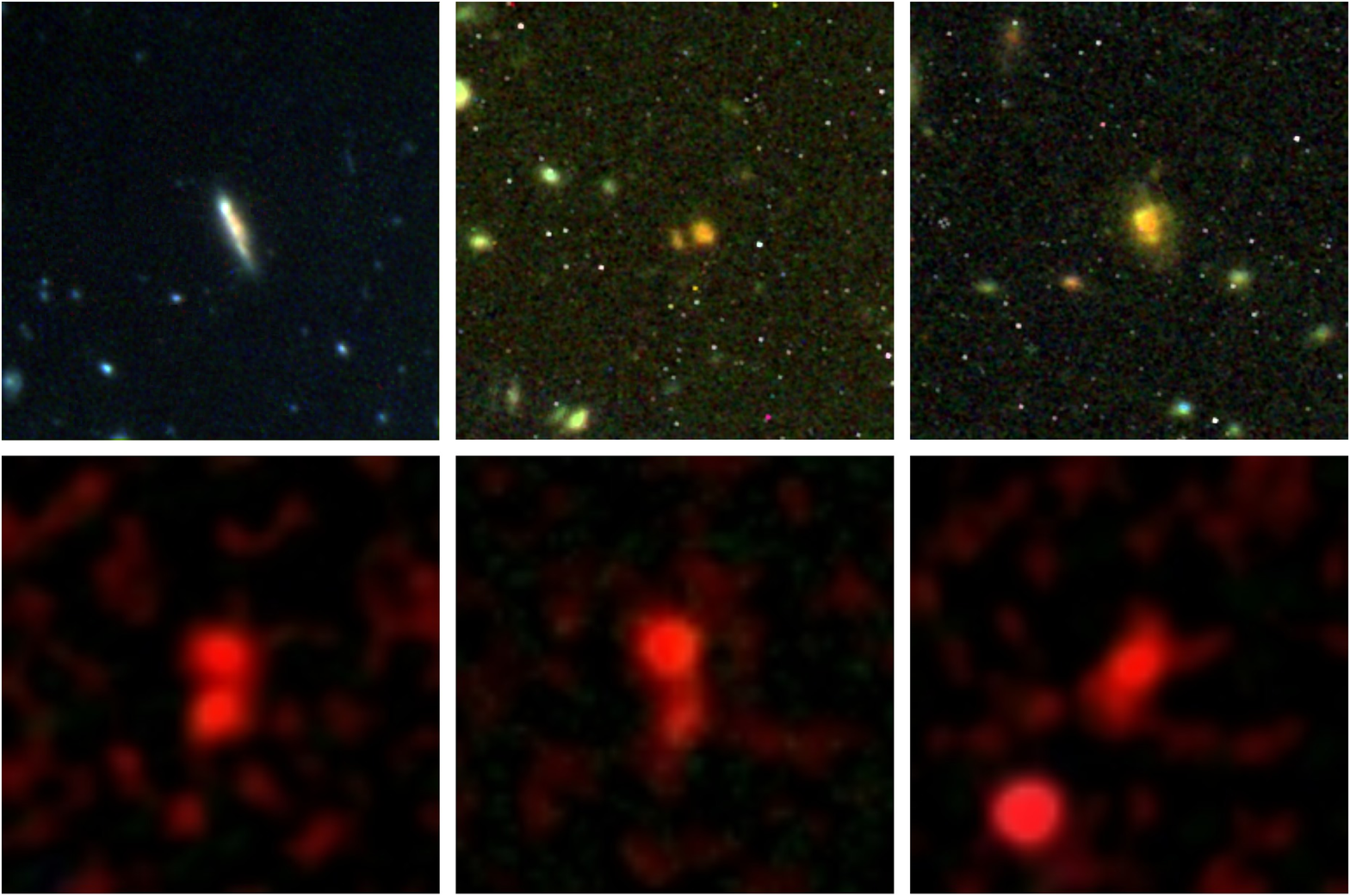}
\noindent{{\bf Fig.~3} Rest-frame optical images of 6 mid-IR-selected
heavily-obscured quasars at $z$$\sim$2 in the ECDF-S region. Top panel show images
obtained using the HST/WFC3 camera using the $Y$, $J$ and $H$
observations of the Ultra-Deep (left image) and GOODS fields. The
bottom images were made combining data in the $R$, $J$, and $K$ bands
obtained from ground-based telescopes, hence with $\sim$10$\times$ worse spatial
resolution than the HST images. All images are
15$''$$\times$15$''$. }
\end{figure*}

\vfill
\eject

\clearpage

\begin{center}
\huge
\textbf{Supporting Online Material}
\bigskip
\normalsize
\end{center}
\baselineskip16pt

\section*{Observations of obscured quasars}

Finding obscured sources is a difficult task. So difficult that in
fact for a while it was suspected that these sources did not
exist. However, now mostly due to extensive multiwavelength surveys a
significant number of these sources have been identified. In this
work, we have compiled observations of obscured quasars selected at
various wavelengths using different techniques, which are described
below.

In the nearby Universe, we can estimate the number of heavily-obscured
quasars from the space density of ULIRGs. In particular, we base our
calculation in the IRAS Revised Bright Galaxy Sample
\cite{sanders03}. This sample includes both obscured and unobscured
quasars, since only an IR selection was done.  A random subsample of
16 sources with $L_{\rm IR}$$>$10$^{12}$$L_\odot$ was observed in
X-rays by {\it Chandra} \cite{iwasawa09}.  Of those, 9 have clear AGN
signatures, namely a flat X-ray spectrum, and hence the fraction of
AGN in this ULIRG sample is at least 56\%. However, for the remaining
7 sources a strong Fe K emission lines is detected in the stacked
X-ray signal. Since detection of Fe K in emission is considered an
unambiguous AGN signature, we can conclude that the fraction of
sources containing an AGN, even in the sample of 7 sources not
dominated by the AGN emission, should be very high. Similarly,
although at higher redshifts, from a sample of 70 $\mu$m-selected
sources in the Cosmic Evolution Survey (COSMOS) field detected with
{\it Spitzer}, J. Kartaltepe et al. \cite{kartaltepe10} conclude that
the fraction of sources with AGN signatures at $L_{\rm
IR}$$>$10$^{12}$$L_\odot$ is higher than 80\% and 100\% for $L_{\rm
IR}$$>$10$^{13}$$L_\odot$. Hence, our assumption that most (if not
all) of the ULIRGs contain an AGN is well justified.

Heavily-obscured quasars can in principle be selected by their X-ray
emission. The effects of obscuration are more important at softer
energies, while harder X-ray photons can in principle escape more
easily.  This is why it is possible to study even Compton-thick
sources at high redshifts using the Chandra bands, which trace
rest-frame energies $>$10 keV at $z$$>$1. There are two types of
Compton-thick quasars based on their X-ray spectra:
transmission-dominated sources, characterized by a photoelectric
cutoff at $E$$\sim$10~keV corresponding to a
$N_H$$\simeq$10$^{24}$~cm$^{-2}$ and in which the intrinsic emission
is visible at higher energies, and reflection-dominated quasars, which
are in general more obscured sources in which the intrinsic emission
at X-ray wavelengths is completely absorbed and only a weak
($\sim$1-2\% of the intrinsic luminosity) reflection component can be
detected.

Using the deep Chandra observations available in the Chandra Deep
Field South (CDF-S), P. Tozzi et al. \cite{Stozzi06} identified 10
Compton-thick candidates (both transmission and reflection dominated)
with intrinsic luminosity $L_X$$>$10$^{44}$~erg/s in the
0.73$<$$z$$<$4.29 range. These sources were used to compute the space
density of Compton-thick AGN in two redshift bins at $z$$\sim$1 and
$z$$\sim$2, as shown in Fig.~1 in the main text.

As most of the absorbed energy is re-radiated at infrared wavelength,
obscured AGN are in general very bright in the mid-IR bands (e.g.,
\citen{treister06}). Using a combination of optical and mid-IR
spectroscopy, together with X-ray data, seven Compton-thick quasar
candidates at $z$$\simeq$2 were identified \cite{Salexander08}. The
space density of Compton-thick quasars was estimated by these authors
from the four sources found in the GOODS-North field to be
$\simeq$0.7-2.5 $\times$10$^{-5}$ Mpc$^{-3}$.

Using a method based only on optical to mid-IR photometry, F. Fiore et
al. \cite{fiore08} showed that the sources with a high mid-IR to
optical flux ratio present at the same time significant X-ray
emission, higher than the expected value for pure star-forming
galaxies and consistent with Compton-thick obscuration levels.  This
selection method was then applied to the sources in the COSMOS field
\cite{Sfiore09}. This provides a space density of Compton-thick quasars
of 4.8$\pm$1.1$\times$10$^{-6}$ Mpc$^{-3}$ in the 1.2$<$$z$$<$2.2
redshift bin. As can be seen in Fig.~1 of the main paper, this value
is in very good agreement with the density derived from X-ray
observations \cite{Stozzi06} for sources at similar redshifts using a
completely independent selection technique.

The same optical to mid-IR selection technique was used
 in the Extended CDF-S (ECDF-S) field \cite{treister09}, which has
very deep {\it Spitzer} and {\it Chandra} imaging over a $\sim$0.3 deg$^2$ region
of the sky. A total of 211 Compton-thick quasar candidates were found
using the selection scheme of F. Fiore et al. \cite{fiore08}. Thanks to the larger
number of sources in this sample, it is possible to measure the
density of heavily-obscured sources in four redshift bins, from
$z$$\sim$1.5 to $z$$\sim$2.9.

\section*{Estimating the ratio of obscured to unobscured quasars from galaxy mergers}

In order to calculate the redshift dependence of the ratio of obscured
to unobscured quasars we start with the following reasonable and
justifiable assumptions. New quasars are ignited by the gas-rich merger of two
massive galaxies. The product of this merger produces a
ULIRG, which harbors at its center a heavily-obscured quasar. After a
time $\Delta$t most of the surrounding dust and gas are blown out by
the radiation pressure of the rapid on-going accretion, after which an
unobscured optically bright quasar is visible. In this scenario, the
ratio of obscured to unobscured quasars as a function of redshift is
simply given by the merger rate of gas-rich galaxies (equation 1) in
the main text of the paper. Here, we describe and evaluate each term
in that equation and its redshift dependence in detail.

One of the fundamental ingredients in this calculation is the evolution of
the merger rate and its dependence on other parameters like e.g., galaxy
mass. This is still a controversial topic. While work based on galaxy morphology
to identify mergers reported no redshift dependence (e.g., \citen{lotz08}),
others based on galaxy pairs (e.g., \citen{kartaltepe07}) found a strong
redshift dependence of the form $\simeq$(1+$z$)$^3$. A compilation
by P. Hopkins et al. \cite{Shopkins09}, which includes the sources in the two samples mentioned
before, concluded that once the differences in median galaxy mass and mass ratio (minor
versus major mergers) are accounted for, the dependence of the merger
rate on redshift is confirmed at a high statistical significance. Following
this work, the major merger rate per galaxy as derived from simulations constrained 
by observations is parametrized as:

\begin{equation}
\frac{d^2{\rm Merger}}{dtdN}=A(M_{\rm Min})(1+z)^{\beta(M_{\rm Min})}
\end{equation}
where 

\[
A(M_{\rm Min})=0.02[1+(\frac{M}{2\times10^{10}M_\odot})^{0.5}] ~\textnormal{Gyr$^{-1}$}
\]
and
\[
\beta(M_{\rm Min})=1.65-0.15\log(\frac{M}{2\times10^{10}M_\odot}).
\]
Uncertainties in these parameters are a factor of $\sim$2 in $A(M_{\rm Min})$ and 0.2 in $\beta(M_{\rm Min})$. 
The minimum stellar mass in our definition of ``massive'' galaxy is a
function of redshift, as shown in Fig.~S1, and can be
parametrized as

\begin{equation}
M_{\rm Min}(z)=5\times10^{11}(1+z)^{-1.5}M_\odot.
\end{equation}
Hence, $A(M)$ ranges from 0.12 Gyrs$^{-1}$ at $z$=0 to 0.06
Gyrs$^{-1}$ at $z$=3, while $\beta(M)$ goes from 1.44 at $z$=0 to 1.58
at $z$=3. The resulting redshift dependence of the merger rate per
galaxy is shown in Fig.~S2.

The space density of galaxies as a function of stellar mass and
redshift was determined observationally recently by D. Marchesini et
al. \cite{Smarchesini09}.  They report that the stellar mass function
is well-fit by a Schechter \cite{schechter76} function up to
$z$$\sim$4 in which the normalization is a function of redshift. To
calculate the merging rates of massive galaxies, we use the limiting
stellar mass given by equation (2). The resulting space density of
massive galaxies as a function of redshift is shown by the blue line
on Fig.~S2. In order to estimate the number of
gas-rich galaxies the extinction-corrected star formation rate density
as a function of redshift is used as a proxy of available gas mass at
each epoch. This is done because direct measurements of gas contents,
especially in high redshift galaxies, are very difficult and hence we
have to rely on secondary indicators. In particular, we use the
measurements obtained from the Great Observatories Origins Deep Survey
(GOODS) observations \cite{Sdahlen07}. A good fit to these data,
normalized to the high-redshift value, is obtained using the following
expression

\begin{equation}
\textnormal {Gas rich}(z)=\left\{ \begin{array}{l l}
0.11(1+z)^{2.0}  & z\leq 2 \\ 
1 & z>2
\end{array}\right..
\end{equation}

This dependence (shown in Fig.~S2) is consistent
with results obtained from hydrodynamical simulations
\cite{hopkins07}.  Finally, the number of unobscured quasars is
obtained directly from observations, in particular from X-ray
luminosity functions \cite{Shasinger05}. Since this selection was made
at soft X-ray energies, this ensures that only unobscured sources are
included. Only sources with $L_X$$>$10$^{44}$~erg/s, corresponding to
$L_{\rm bol}$$>$10$^{45}$~erg/s, were considered.  If instead an
optical luminosity function (e.g., \citen{richards06}) is used,
assuming in this case an equivalent magnitude cut of $M_i$$<$-23
corresponding to a similar luminosity threshold, the results are
consistent. As can be seen in Fig.~S2, the quasar
number density has a strong peak at $z$$\sim$1.5, followed by a
shallow decline towards higher redshifts.

The redshift dependence of each of the components in our calculation
is shown in Fig.~S2. Both the number of unobscured
quasars and massive galaxies show steep declines from $z$$\sim$1 to
$z$=0. Meanwhile, the merger rate increases steadily by a factor of
$\sim$3 from $z$=0 to $z$=3.

\begin{figure*}[h!]
\begin{center}
\includegraphics[scale=0.75]{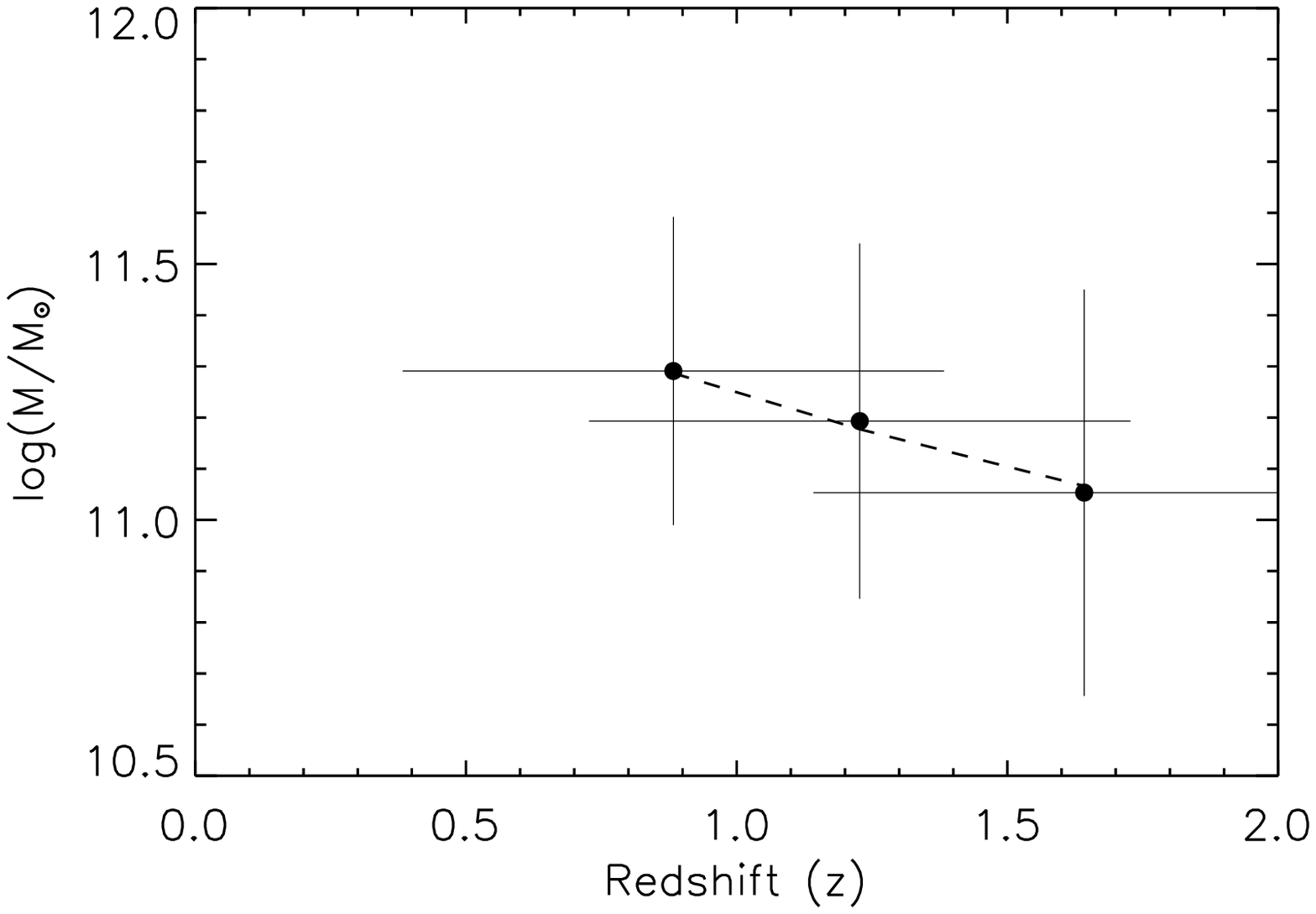}
\vspace{-0.6cm}
\noindent{\\{\bf Fig.~S1.} Median stellar mass as a function of redshift for the host
galaxies of the ULIRGs (L$_{\rm IR}$$>$10$^{12}$L$_\odot$) in the COSMOS
field. Stellar masses were derived by performing spectral fitting of
the observed-frame optical and near-IR photometric data
\cite{ilbert09}. Vertical error bars were obtained from the
dispersion in each bin, while the horizontal error bar show the width
of each bin.  The dashed line shows the best-fitting power-law,
5$\times$10$^{11}$(1+$z$)$^{-1.5}$ M$_\odot$, which is used to compute
the number of galaxies hosting ULIRGs as a function of redshift.}
\label{mass_red}
\end{center}
\end{figure*}

\begin{figure*}[h!]
\begin{center}
\includegraphics[scale=0.6]{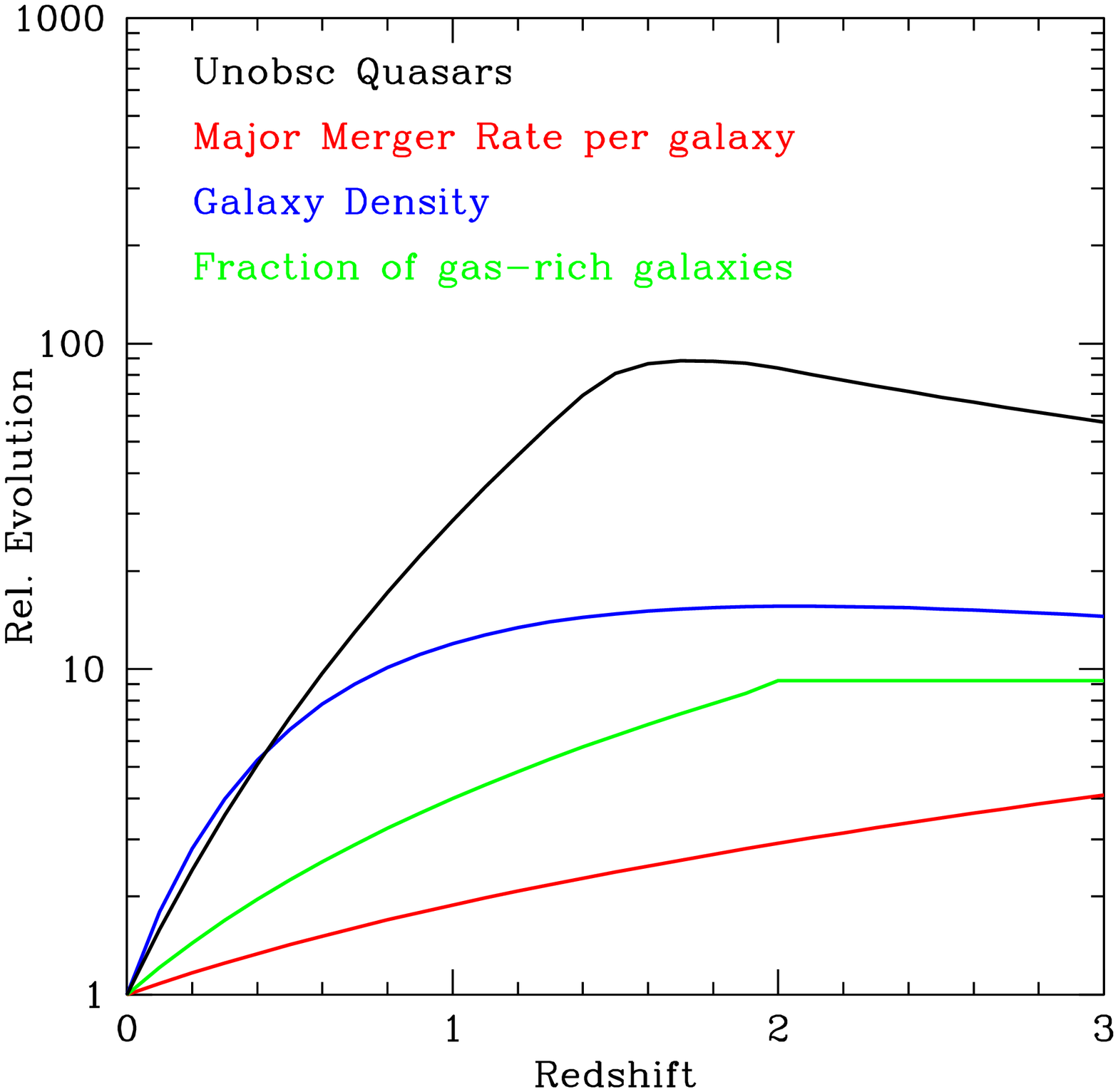}
\noindent{\\{\bf Fig.~S2.} Evolution of the different ingredients used in our
calculation of the expected ratio of obscured to unobscured quasars as
a function of redshift. All curves were normalized to their $z$=0
value. {\it Black line} shows the evolution of the density of
unobscured quasars, while the redshift dependence of the major merger
frequency per galaxy, density of massive galaxies and fraction of
gas-rich galaxies is shown by the {\it red}, {\it blue} and {\it
green} lines.  }
\label{ratio_mergers}
\end{center}
\end{figure*}


\begin{thebibliography}{10}
\bibitem{note1} By quasar we refer here to the most luminous members
of the Active Galactic Nuclei (AGN) family, typically $M_{\rm
B}$$<$-23, $L_X$$>$10$^{44}$~erg/s or $L_{\rm bol}$$>$10$^{45}$~erg/s.
\bibitem{note2}
By heavily-obscured we refer here to
sources with neutral hydrogen column densities
$N_{\rm H}$$>$10$^{23}$cm$^{-2}$, enough to hide most of the soft X-rays to
optical quasar signatures.
\bibitem{note3}
This integrated value was obtained assuming
an accretion efficiency of 0.1 and is in very good agreement with
observational results \cite{marconi04}, even after incorporating this
additional accretion.
\end{thebibliography}
\end{document}